\documentclass[aps,prc,groupedaddress,showpacs,manuscript]{revtex4}

\usepackage{graphicx}

\begin{document}

\title{Difficulties in probing density dependent symmetry potential with the HBT interferometry}

\author {LI QingFeng \footnote{Corresponding author (e-mail: liqf@hutc.zj.cn)}
\& SHEN CaiWan }
\address{
School of Science, Huzhou Teachers College, Huzhou 313000,
China\\
 }


\begin{abstract}
Based on the updated UrQMD transport model, the effect of the
symmetry potential energy on the two-nucleon HBT correlation is
investigated with the help of the coalescence program for
constructing clusters, and the CRAB analyzing program of the
two-particle HBT correlation. An obvious non-linear dependence of
the neutron-proton (or neutron-neutron) HBT correlation function
($C_{np,nn}$) at small relative momenta on the stiffness factor
$\gamma$ of the symmetry potential energy is found: when $\gamma
\lesssim 0.8$, the $C_{np,nn}$ increases rapidly with increasing
$\gamma$, while it starts to saturate if $\gamma \gtrsim 0.8$. It is
also found that both the symmetry potential energy at low densities
and the conditions of constructing clusters at the late stage of the
whole process influence the two-nucleon HBT correlation with the
same power.

\hspace{2cm}

\end{abstract}

 \maketitle

In recent years, the isospin dependent equation of state (EoS) of
both infinite and finite nuclear matter has been investigated in
both deeper and broader range. Theoretically, a large amount of
probes have been brought out to be sensitively affected by the
isovector part of the EoS, i.e., the symmetry energy term. Based on
different theories, the symmetry energy has been parametrized in
different ways (please see \cite{Bro00,Baran:2004ih,Li:2008gp} for
details). Among these parametrizations, the
$E_{sym}=S_{0}u^{\gamma}$ is often used ($S_{0}$ being the symmetry
energy at the normal density, and $u=\rho/\rho_0$ the reduced
density, while $\gamma$ is the stiffness factor of the density
dependence of the symmetry energy). Especially, the density
dependence of the symmetry potential energy is paid much more
attention since the uncertainty of the symmetry energy mainly comes
from this term. Recently, with some comparisons between the results
of experimental data and dynamical model IBUU, the uncertainty in
the density dependence of the symmetry energy at subnormal densities
is reduced into the range of $\gamma\sim 0.7-1.1$, depending on the
probable medium modifications of the nucleon transport
\cite{Chen:2004si,Li:2005jy}. Nevertheless, recently a quite soft
symmetry energy (the corresponding $\gamma$ factor is only about
$0.3$) was implied experimentally at very low nuclear densities
($\rho_N$ is $0.01-0.05$ times normal density)
\cite{Kowalski:2006ju}. Furthermore, in the most recent analyses by
Tsang's group \cite{Tsang:2008fd}, although they found that several
different observables can provide consistent constraints on the
density dependence of the symmetry energy, the constraints are still
relatively loose. To bring forward more sensitive probes for
symmetry energy at subnormal densities is still urgent and
prerequisite. However, we think that one should firstly check more
carefully the difficulties of detecting the existing observables for
forthcoming experiments. Therefore, the currently crucial question
becomes: Can the sensitive observable suggested by theoretical
physicists be {\it experimentally} taken as a sensitive candidate
for detecting the symmetry energy at subnormal densities? In the
previous work \cite{Li:2008fn} this question has been concerned for
the free neutron/proton ($n/p$) ratio as well as the
$\Delta^0/\Delta^{++}$ ratio.

In this paper, we would like to continue this topic to further check
the two-nucleon correlation, which is based on the
Hanbury-Brown-Twiss interferometry (HBT)
\cite{HBT54,Goldhaber60,Bauer:1993wq} technique. It was pointed out
that it is a good candidate to probe the symmetry energy since it is
quite unsensitive to the iso-scalar part of EoS as well as the
in-medium NN cross section \cite{Chen:2003wp}. In their initiated
investigations, the isospin effect is found to occur only in the
late stage of the heavy ion collisions (HICs) where the nuclear
medium becomes dilute. Therefore, the two-nucleon correlation has
the advantage in probing the density dependence of the symmetry
potential at low densities. It was further claimed that the
sensitivity of the correlation function of the nucleon pair to the
density dependence of the symmetry energy is reduced after
considering a (isospin-dependent) momentum dependent nuclear
potential \cite{Chen:2004kj}. However, in their studies, only
several parameter sets of the stiffness of the symmetry energy were
employed. It is believed to be enough in the studies of the density
dependent symmetry energy by adopting the conventional
single-particle observables since they behave monotonously with the
increase of the stiffness factor $\gamma$. But, is that still true
in the symmetry-energy dependence of the two-nucleon HBT
correlation? Here we would like to vary the $\gamma$ factor with a
larger range to see the variance of the correlation function.

Besides the symmetry potential energy, the freeze-out conditions of
clusters should also heavily influence the final two-nucleon
correlation function. In the transport model calculations, when the
transport is stopped at a certain time, such as $150$ fm$/c$ or so,
one cluster constructing program is then used. Generally speaking,
two methods (i.e., BUU-like and QMD-like) for constructing clusters
exist in the transport model calculations \cite{Li:2008fn}. In our
UrQMD calculations below, the coalescence model, which is often used
for analyzing QMD-like model outputs, is employed, in which the
nucleons with relative momenta smaller than $P_0$ and relative
distances smaller than $R_0$ are considered to belong to one
cluster. It is comparable to the BUU-like coordinate density cut
$\rho_c$ when we only adopt the relative distance $R_0$ but not the
relative momentum $P_0$ in the QMD-like analysis. The freeze-out
condition certainly influences the two-nucleon correlation function
since it reflects the spatial (and momentum) separation of two
nucleons. It is also interesting to see if the small uncertainty in
the multiplicity of the observed particle changes the HBT
correlation function accordingly.

In this paper, we use the recently updated UrQMD transport model
\cite{Li:2006wc,Li:2007yd} for studies of intermediate energy HICs
to investigate the effects of the density dependent symmetry
potential energy and the freeze-out conditions on the neutron-proton
and the neutron-neutron HBT correlations. The proton-proton
correlation will not be shown in this paper since it has similar
isospin-dependent effect to the neutron-proton and neutron-neutron
ones, although the largest effect is shown at the relative momentum
of pair $q\sim 20$ MeV$/c$ but not at $q \lesssim 10$ MeV$/c$. As a
default, $P_0$ and $R_0$ used in the coalescence model are set to be
$0.3$ GeV$/c$ and $3.5$ fm, respectively (except otherwise stated).
The central collisions ($<11\%$ of total cross section $\sigma_T$)
of neutron-rich intermediate-mass and heavy systems
$^{52}$Ca$+^{48}$Ca and $^{197}$Au$+^{197}$Au are adopted in the
UrQMD calculations. As stated above, the correlation function of the
nucleon pair from HICs with the super-intermediate-mass system at
the intermediate energy is claimed to be unsensitive to the
isospin-scalar part of EoS and the NN cross section
\cite{Chen:2003wp,Li:2006ez}. It is also reported that by using the
imaging method the shape of the source of the two-proton pairs from
the HICs with medium sized system at beam energies around $100$A MeV
is sensitive to the medium modification of the NN cross section
\cite{Verde:2003cx}. However, the corresponding HBT correlation
function is seen to be less and less affected by the medium
modification with the increase of the beam energy. In this work, we
focus our investigation on the correlation functions at beam
energies $100$A $\sim 1000$A MeV. And, we randomly adopt several
parameter sets of EoS used in previous works \cite{Li:2005gf}, for
checking. Those are, S-EoS, SM-EoS, and HM-EoS, without considering
the medium modifications on NN cross sections.

To analyze the two-nucleon HBT interferometry, the Correlation After
Burner (CRAB v3.0$\beta$) program is employed
\cite{HBT54,Goldhaber60,Pratthome}. The correlation function is
expressed as $C(P,q)=B/A$, where $B=\int d^4x_1 d^4x_2 g(x_1,{\bf
P}/2) g(x_2,{\bf P}/2) |\phi({\bf q}, {\bf r})|^2$ is the inclusive
two-particle emission probability in the same event, $A={\int d^4x_1
g(x_1,{\bf P}/2)} {\int d^4x_2 g(x_2,{\bf P}/2)}$ is the product of
two single-particle emission probability in different events. The
squared relative two-particle wave function $|\phi|^2$ serves as a
weight, while $g(x,{\bf P}/2)$ is the probability for emitting a
particle with momentum ${\bf P}/2$ from the space-time point $x =
({\bf r}, t)$. The ${\bf P}={\bf p}_1+{\bf p}_2$ and the ${\bf
q}=({\bf p}_1-{\bf p}_2)/2$ are the total and relative momenta of
the particle pair, respectively.

\begin{figure}
\includegraphics[angle=0,width=0.6\textwidth]{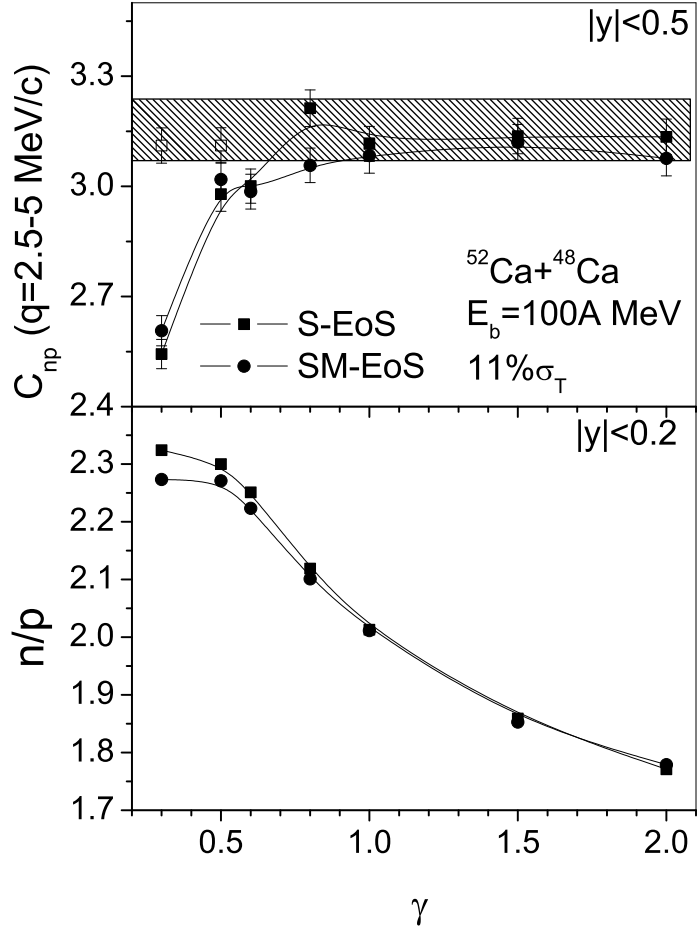}
\caption{Neutron-proton correlation function ($C_{np}$) within the
relative momentum $q=2.5-5$ MeV$/c$ and the rapidity region
$|y_{cm}<0.5|$ (top plot), and free neutron-proton ratio in the
rapidity region $|y_{cm}<0.2|$ (bottom plot) as a function of the
stiffness factor ($\gamma$) of the symmetry potential energy. The
S-EoS and SM-EoS are adopted for $^{52}$Ca$+^{48}$Ca central
collisions ($<11\%$ of total cross section $\sigma_T$) at $E_b=100$A
MeV. The marked area illustrates the value $C_{np}=3.15\pm0.08$. The
open squares in the top plot show the $C_{np}$ values when the
symmetry potential energy is switched off at the reduced densities
$u<0.2$.} \label{fig1}
\end{figure}

Figure\ \ref{fig1} shows the neutron-proton correlation function
($C_{np}$) within the relative momentum bin $q=2.5-5$ MeV$/c$ and
the rapidity region $|y_{cm}<0.5|$
($y_{cm}=\frac{1}{2}\log(\frac{E_{cm}+p_{//}}{E_{cm}-p_{//}})$,
$E_{cm}$ and $p_{//}$ are the energy and longitudinal momentum of
the nucleon in the center-of-mass system) (top plot), and the free
$n/p$ ratio in the rapidity region $|y_{cm}<0.2|$ (bottom plot) as a
function of the stiffness factor $\gamma$ of the symmetry potential
energy. The $^{52}$Ca$+^{48}$Ca central collisions ($<11\%$ of total
cross section $\sigma_T$) at $E_b=100$A MeV are adopted. It is known
that the $n/p$ ratio of free nucleons is obviously rapidity
dependent. But conclusions of the two-nucleon HBT correlation drawn
in this paper is not changed by the small rapidity-cut difference
used in both observables \cite{Li:2006gb}. The S-EoS and the SM-EoS
are selected for comparison in each plot. It is seen that the
iso-scalar part of the EoS only affects slightly the sensitivity of
both the $C_{np}$ and the $n/p$ ratio to the stiffness of the
symmetry potential energy. Thus, in the following calculations, we
will randomly use various EoS. It is also known that the $n/p$ value
decreases monotonously with increasing $\gamma$ factor, which is
again shown in the bottom plot of this figure. It is interesting to
see that, however, the $C_{np}$ moves up rapidly with increasing
$\gamma$ when $\gamma$ is less than $\sim 0.8$, and then saturates
with the further increase of the $\gamma$ value, the saturation
value is $C_{np}=3.15\pm0.08$, which is shown in a marked area. It
is known that the symmetry potential plays obvious role on the
two-nucleon HBT correlation at the late stage ($t>50$ fm$/c$) of the
whole process \cite{Chen:2003wp}, where the density surrounding the
emitted protons is very small. Therefore, we further calculate the
$C_{np}$ with $\gamma=0.3$ and $0.5$ (open squares) but without
considering the symmetry potential energy at very low densities (the
reduced density $u=\rho/\rho_0<0.2$). It is obvious that the
$C_{np}$ values move back to the marked area. We know that the
stiffer the symmetry energy is, the smaller the value of the
symmetry energy at subnormal density becomes. At densities $u<0.2$
the stiff symmetry energy is too weak to influence the two-particle
correlation, while it plays strong role with a very soft symmetry
energy \cite{Li:2008fn}. It is concluded that, firstly, the
correlation function can hardly distinguish the stiffness of the
symmetry energy when $\gamma>0.8$. Secondly, the soft symmetry
energy plays more obvious role on the neutron-proton correlation
function at the late stage. Therefore, the neutron-proton
correlation function is suitable to explore the symmetry energy in
the dilute nuclear medium, especially for a very soft
symmetry-energy assumption.

\begin{figure}
\includegraphics[angle=0,width=0.8\textwidth]{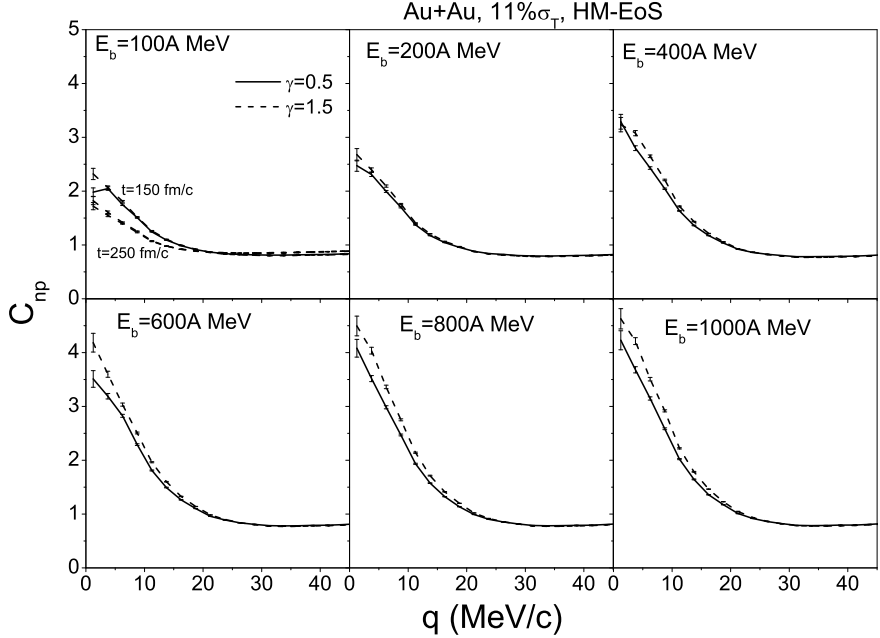}
\caption{Neutron-proton correlation functions in central Au+Au
collisions at $E_b=100$A, $200$A, $400$A, $600$A, $800$A, and
$1000$A MeV. The UrQMD calculations are stopped at $150$ fm$/c$. At
$E_b=100$A MeV, the results at $250$ fm$/c$ are also shown. The
HM-EoS and the symmetry energies with $\gamma=0.5$ and $1.5$ are
adopted in calculations.} \label{fig2}
\end{figure}

At beam energy around $100$A MeV, it is found that the isospin
effect on the correlation function is reduced in heavy-system HICs
such as Sn+Sn. It is also confirmed in our calculations with the
Au+Au reaction which is shown in the upper-left plot of Figure\
\ref{fig2}, even when a different stop time $250$ fm$/c$ is chosen.
Further, we calculate the $C_{np}$ values at higher beam energies,
i.e., at $E_b=200$A, $400$A, $600$A, $800$A, and $1000$A MeV, which
are shown in other plots of Figure\ \ref{fig2}. It is interesting to
see that with the increase of the beam energy, the effect of the
density dependence of symmetry energy becomes visible. It might be
understandable when considering the fact that with the increase of
beam energy, more unstable light fragments emit and decay with a
longer time and in the dilute medium so as to be influenced by the
soft symmetry energy at the late stage.

\begin{figure}
\includegraphics[angle=0,width=0.7\textwidth]{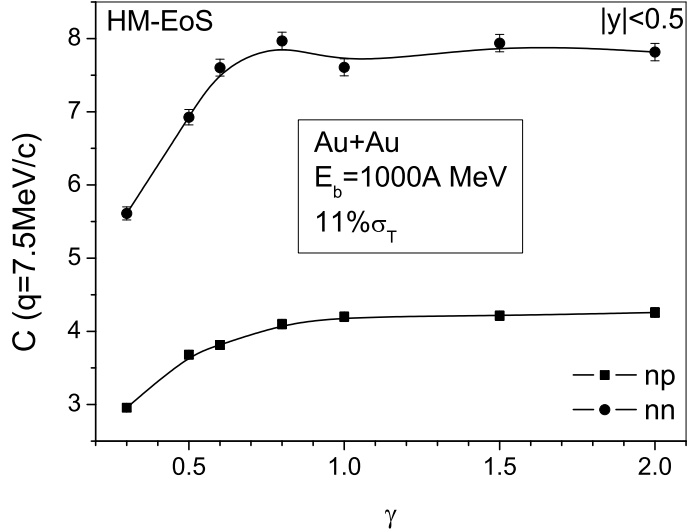}
\caption{Neutron-proton (line with squares) and neutron-neutron
(line with circles) correlation functions within the relative
momentum $q=2.5-5$ MeV$/c$ and the rapidity region $|y_{cm}<0.5|$ as
a function of the stiffness factor ($\gamma$) of the symmetry
potential energy. The HM-EoS is adopted for Au+Au central collisions
at $E_b=1000$A MeV. } \label{fig3}
\end{figure}

For the neutron-neutron correlation, the phenomena shown in Figs.\
\ref{fig1} and \ref{fig2} are even more obvious. In Figure\
\ref{fig3} we show the $\gamma$-dependence of the correlation
functions of both neutron-proton and neutron-neutron pairs from
central Au+Au collisions at $E_b=1000$A MeV. The strong non-linear
dependence of the neutron-neutron correlation function on the
$\gamma$ factor of the symmetry potential energy is due to the long
time evolution of the neutron-rich unstable light fragments at the
late stage.

\begin{figure}
\includegraphics[angle=0,width=0.7\textwidth]{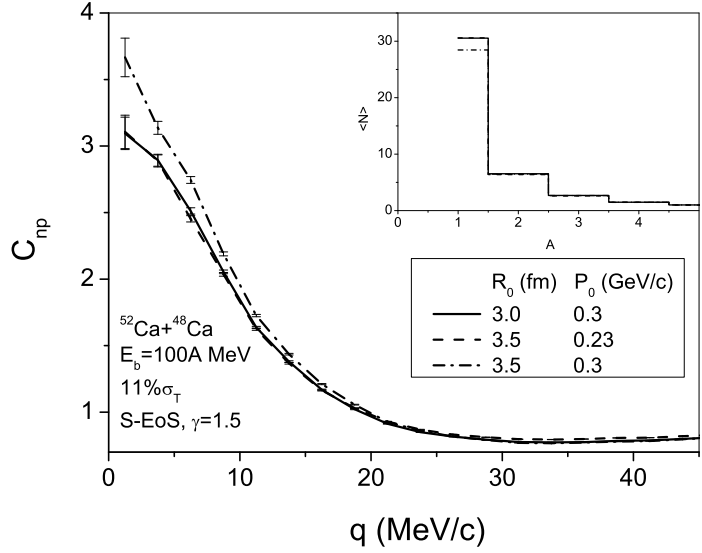}
\caption{Neutron-proton correlation functions from central
$^{52}$Ca$+^{48}$Ca collisions at $E_b=100$A MeV are illustrated
with different relative momenta $P_0$ and relative distances $R_0$
in the coalescence model, and in the top-right plot the
multiplicities of light clusters ($A<5$) are shown, correspondingly.
The S-EoS and the symmetry potential energy with $\gamma=1.5$ are
used. } \label{fig4}
\end{figure}

In order to deeper understand the isospin effect on the correlation
function, it is necessary to check the influence of the freeze-out
conditions on the HBT correlator. Figure\\ \ref{fig4} illustrates
the neutron-proton correlator as a function of the relative momentum
$q$ of the nucleon pair. Central $^{52}$Ca$+^{48}$Ca collisions at
$E_b=100$A MeV are calculated with UrQMD and sets of relative
momentum ($P_0$) and relative distance ($R_0$) are chosen for the
following cluster construction in the coalescence model. The final
correlators by the CRAB analyzing program are shown with different
lines in the figure. The multiplicities of the light clusters
($A<5$) with different ($P_0$,$R_0$) sets are shown in the top-right
plot. The stiffness of the symmetry potential energy $\gamma=1.5$ is
chosen in current calculations. When the values of ($P_0$, $R_0$)
are adjusted from ($0.3$ GeV$/c$, $3.0$ fm) to ($0.23$ GeV$/c$,
$3.5$ fm), the multiplicities of clusters keep almost unchanged, and
so does the correlator. However, when the values of ($P_0$, $R_0$)
are changed from ($0.3$ GeV$/c$, $3.0$ fm) to ($0.3$ GeV$/c$, $3.5$
fm), the multiplicities of free nucleons alter slightly, and so does
the correlator. This vivid phenomenon reveals that the freeze-out
condition influences the value of the correlator visibly and should
be paid more attention. In order to constrain the density dependence
of the symmetry energy by using the two-nucleon HBT correlation, the
uncertainty of the multiplicity of nucleons, with whatever
experimental cuts, should be largely reduced firstly.

To summarize, based on the updated UrQMD transport model, the
coalescence program for constructing clusters, and the CRAB
analyzing program of the two-particle HBT correlation, a more
detailed analysis of the effect of the symmetry potential energy on
the two-nucleon HBT correlation is investigated for HICs with super
medium-sized system at intermediate energies. It is found that, for
HICs with medium-sized neutron-rich system at $100$A MeV, there
exists a non-linear dependence of the neutron-proton (or
neutron-neutron) HBT correlation function ($C_{np,nn}$) at small
relative momenta $q$ on the stiffness factor $\gamma$ of the
symmetry potential energy: when $\gamma \lesssim 0.8$, the
$C_{np,nn}$ at small $q$ increases rapidly with increasing $\gamma$,
while it starts to saturate if $\gamma \gtrsim 0.8$. This phenomenon
is also seen for the HICs with the heavy system at higher beam
energies. At $E_b=1000$A MeV, the non-linear dependence of the
$C_{nn}$ is even more obvious than the neutron-proton case. With the
upgrading of the LAND facility at GSI to the ``newLAND'', it is
expected to detect the stiffness of the symmetry energy with more
neutron-related observables in the near future.

In order to understand the non-linear dependence of the HBT
correlator on the $\gamma$ factor, the different freeze-out
conditions, such as the stop times and the ($P_0$,$R_0$) parameter
sets for the coalescence model, are taken into account. It is found
that both the symmetry potential energy at low densities and
conditions of constructing clusters at the late stage of the whole
process influence the two-nucleon HBT correlation with the same
power and should be studied more carefully.

\section*{Acknowledgments}
We would like to thank S. Pratt for providing the CRAB program and
En-Guang Zhao for helpful discussions and encouragements. We
acknowledge support by the Frankfurt Center for Scientific Computing
(CSC). The work is supported in part by the key project of the
Ministry of Education of China under grant No. 209053 and the
National Natural Science Foundation of China under grant No.
10675046.


\newpage
\begin{thebibliography}{99}

\bibitem{Bro00}
Brown B A. Neutron Radii in Nuclei and the Neutron Equation of
State. Phys Rev Lett, 2000, 85:5296

\bibitem{Baran:2004ih}
  Baran V, Colonna M, Greco V and Toro M Di.
  Reaction Dynamics with Exotic Beams.
  Phys Rept, 2005, 410:335


\bibitem{Li:2008gp}
  Li B A, Chen L W and Ko C M.
  Recent Progress and New Challenges in Isospin Physics with Heavy-Ion
  Reactions.
  Phys Rept, 2008,  464:113



\bibitem{Chen:2004si}
  Chen L W, Ko C M and Li B A.
  Studying stiffness of nuclear symmetry energy through isospin diffusion in
  heavy-ion collisions.
  Phys Rev Lett, 2005, 94:032701

\bibitem{Li:2005jy}
  Li B A and Chen L W.
  Nucleon-nucleon cross sections in neutron-rich matter and isospin transport
  in heavy-ion reactions at intermediate energies.
  Phys Rev C, 2005, 72:064611

\bibitem{Kowalski:2006ju}
  Kowalski S, et al.
  Experimental determination of the symmetry energy of a low density nuclear
  gas.
  Phys Rev  C, 2007, 75:014601

\bibitem{Tsang:2008fd}
  Tsang M B, Zhang Y, Danielewicz P, Famiano M, Li Z, Lynch W G and Steiner A W.
  Constraints on the density dependence of the symmetry energy.
  Phys Rev Lett, 2009, 102:122701



\bibitem{Li:2008fn}
  Li Q.
  The influence of reconstruction criteria on the sensitive probes of the
  symmetry potential.
  Mod Phys Lett A, 2009, 24:41

\bibitem{HBT54}Hanbury-Brown R and Twiss R Q. Philos Mag, 1954, 45:663; Nature (London), 1956, 178:1046

\bibitem{Goldhaber60}
Goldhaber G, et al. Phys Rev, 1960, 120:300

\bibitem{Bauer:1993wq}
  Bauer W, Gelbke C K and Pratt S.
  Hadronic interferometry in heavy ion collisions.
  Ann Rev Nucl Part Sci, 1992, 42:77

\bibitem{Chen:2003wp}
  Chen L W, Greco V, Ko C M and Li B A.
  Isospin effects on two-nucleon correlation functions in heavy-ion
  collisions at intermediate energies.
  Phys Rev C, 2003, 68:014605


\bibitem{Chen:2004kj}
  Chen L W, Ko C M and Li B A.
  Effects of momentum-dependent nuclear potential on two-nucleon correlation
  functions and light cluster production in intermediate energy heavy-ion
  collisions.
  Phys Rev C, 2004, 69:054606



\bibitem{Li:2006wc}
  Li Q, Li Z and St\"ocker H.
  Probing the symmetry energy and the degree of isospin equilibrium.
  Phys Rev C, 2006, 73:051601


\bibitem{Li:2007yd}
  Li Q, Bleicher M and St\"ocker H.
  The effect of pre-formed hadron potentials on the dynamics of heavy ion
  collisions and the HBT puzzle.
  Phys Lett B, 2008, 659:525


\bibitem{Li:2006ez}
  Li Q, Li Z, Soff S, Bleicher M and St\"ocker H.
  Medium modifications of the nucleon-nucleon elastic cross section in
  neutron-rich intermediate energy HICs.
  J Phys G: Nucl Part Phys, 2006, 32:407

\bibitem{Verde:2003cx}
  Verde G, Danielewicz P, Lynch W G, Brown D A, Gelbke C K and Tsang M
  B.
  Probing Transport Theories via Two-Proton Source Imaging.
  Phys Rev C, 2003, 67:034606


\bibitem{Li:2005gf}
  Li Q, Li Z, Soff S, Bleicher M and St\"ocker H.
  Probing the equation of state with pions.
  J Phys G: Nucl Part Phys, 2006, 32:151




\bibitem{Pratthome}
Pratt S. CRAB version 3. http://www.nscl.msu.edu
/$\sim$pratt/freecodes/crab/home.html.



\bibitem{Li:2006gb}
  Li Q, Bleicher M, Zhu X and St\"ocker H.
  Transport model analysis of the transverse momentum and rapidity dependence
  of pion interferometry at SPS energies.
  J Phys G: Nucl Part Phys, 2007, 33:537





\end{thebibliography}
\end{document}